\documentclass[a4paper,11pt]{article}
\pdfoutput=1 

\usepackage{aas_macros}
\usepackage{jcappub} 

\usepackage[T1]{fontenc} 
\usepackage[dvipsnames]{xcolor}

\newcommand{\bu}{\textbf{u}}

\newcommand{\bx}{\textbf{x}}
\newcommand{\bk}{\textbf{k}}

\newcommand{\bv}{\textbf{v}}

\newcommand{\avg}[1]{\ensuremath{\left\langle #1 \right\rangle}}

\def\kMpc{\, h \, {\rm Mpc}^{-1}}

\def\hn{\hat{n}}

\newcommand{\nbar}{\bar{n}}

\title{Can Fingers of God be Resummed?}

\author[a \dagger]{Shi-Fan Chen}
\author[b]{Alejandro Aviles}

\affiliation[a]{Department of Physics, Columbia University, New York, NY, USA 10027}
\affiliation[\dagger]{NASA Hubble Fellowship Program, Einstein Fellow}
\affiliation[b]{Instituto de Ciencias F\'{\i}sicas, Universidad Nacional Aut\'onoma de M\'exico, Av. Universidad s/n, Cuernavaca, Morelos, C.~P.~62210, M\'exico}

\emailAdd{sc5888@columbia.edu}
\emailAdd{aviles@icf.unam.mx}

\abstract{Fingers of God (FoGs), Doppler shifts due to the small-scale virial motions of galaxies, are one of the most significant challenges for analytic models of galaxy clustering in redshift surveys. In the effective field theory approach, the characteristic scale of these shifts introduces a new, large length scale, limiting the reach of the theory. This has motivated many phenomenological models of FoGs in the literature where their effect is approximated by a single damping function of the line-of-sight wavenumber $\lambda=k\mu$, effectively resumming powers of $\lambda$ due to single-point contractions of small-scale velocities. The goal of this work is to investigate whether such a resummation can be rigorously constructed. We show that common damping models can be understood as the mean-field limit of the single-particle characteristic function (SPCF) of FoG velocities, but that the environmental dependence of the SPCF generates equally large corrections that break the simple damping form. We develop the long-wavelength theory of the SPCF, showing that it admits an expansion in anisotropic bias operators closely related to those derived for selection effects, but with $\lambda$-dependent coefficients satisfying selection rules. We estimate these coefficients using the halo model for DESI-like galaxies, showing that they make significant contributions at the level of the 2- and 3-point functions of galaxies. To answer the titular question: potentially, but at the cost of significantly expanding the operator basis and promoting free coefficients to free functions.}

\begin{document}
\maketitle
\flushbottom

\section{Introduction}

Perturbation theory provides an accurate description of large-scale structure well into the quasi-linear regime. Depending on redshift and tracer, perturbative predictions can remain useful up to wavenumbers of order $k_\text{nl} \sim 0.3\,h\, \text{Mpc}^{-1}$. However, the situation is more restrictive in redshift space. Peculiar velocities enter the mapping from real to redshift space as displacements along the line of sight, allowing motions generated on small physical scales to affect clustering at much larger apparent separations. A particularly severe version of this scale mixing is due to the virial motions of galaxies inside halos: defining a halo as a region with mean interior density \(\Delta\) times the critical density, a characteristic velocity \(V_\Delta=(GM_\Delta/R_\Delta)^{1/2}\) produces a comoving redshift-space displacement
\begin{equation}
   \sigma_{u,\Delta}(M) \equiv \frac{V_\Delta(M)}{aH(a)} =\sqrt{\frac{\Delta}{2}}\,R_h(M),
   \label{eqn:sigmaM}
\end{equation}
where $R_h(M)=R_\Delta/a$ is the comoving halo radius \cite{Cooray2002}. For $\Delta=200$, this gives $\sigma_{u}=10R_h$, showing that intrahalo velocities can generate apparent redshift-space separations much larger than the physical scale on which those velocities are produced. The resulting ``Fingers-of-God'' (FoG) distortions \cite{Jackson1972} can therefore become relevant on scales where the real-space density field and the long-wavelength velocity field are still accurately described by perturbation theory. This difficulty comes from the redshift-space mapping itself, which brings short-scale velocity physics into apparently larger scales, such that redshift-space density fields admit a perturbative expansion over a smaller range of scales than do those in real space. The presence of FoGs in redshift-space clustering therefore leads to an effective expansion in two scales $k\sigma_u$ and $k/k_{\rm nl}$, with terms that scale with the former becoming non-perturbative faster than the latter.

Since the structure of the nonlinear redshift-space mapping is well-understood and the astrophysical origin of the anomalously large FoG scale approximately known, an attractive possibility is to reorganize the perturbative expansion in redshift-space by keeping the long-wavelength fields expanded to a given order while resumming particular diagrams related to the FoG effect \cite{Scoccimarro2004}.
The primary physical effect such re-organizations seek to capture is the erasure of apparent line-of-sight structure by large, stochastic shifts due to FoGs \cite{Peacock1994,Taruya2010}.
One set of phenomenological models, based in configuration-space, are the so-called streaming models, which relate real- and redshift-space clustering through the probability distribution of velocity-induced line-of-sight shifts \cite{Fisher1995,ReidWhite2011,Vlah2016,Bianchi2015}. Even more common are models in Fourier space introducing effective damping functions along the line of sight encoding unresolved FoG velocities \cite{DavisPeebles1983,Park1994,Cole1995,HattonCole1998,Scoccimarro1999}. By necessity, these effective dampings are degenerate with counterterms in the EFT to leading order; however, beyond the perturbative reach of the redshift-space EFT, dampings can absorb velocity contributions that are not captured by a finite-order perturbative expansion, particularly where the damping function is not well captured by a Taylor expansion, and can improve the agreement with the clustering statistics such as the 2-point power-spectrum and 3-point bispectrum multipoles over a finite range of scales. This strategy has been widely adopted in the literature, with simple, 1-parameter functional forms for the effective damping like Gaussians \cite{Peacock1994} and Lorentzians \cite{Sheth1996}. More recently, it has motivated the VDG form \cite{Taruya2010,Bianchi2016,Sanchez2017,Eggemeier2025}, as well as hybrid  models in which explicit line-of-sight damping is incorporated into joint power-spectrum and bispectrum modeling \cite{Bansal2026}.

Empirical agreement notwithstanding, it is not immediately clear that the small-scale velocity statistics of galaxies should be characterized by single-parameter distributions or damping functions and, more generally, whether FoG contributions to galaxy clustering can be brought under theoretical control and appropriately resummed. For example, central and satellite galaxies sample different regions of halo phase space, have different velocity dispersions, and enter a given tracer sample with different relative abundances. Their mixture therefore introduces tracer dependence already at the level of the one-point velocity distribution, hard to square with a universal, 1-parameter form \cite{Okumura2015,Hand2017}. Moreover, the velocity statistics of galaxies and halos are known to depend on their local environments, which are not captured in effective damping models \cite{White2001,Seljak2001,Cooray2002,Tinker2007,Bianchi2015}.

In this work, we will seek to answer the more general question of whether a well-posed resummation can be constructed first, without specifying to particular functional forms for the damping, which we leave for more detailed works utilizing numerical simulations and observational data. Specifically, we wish to factorize and resum contributions due only to stochastic FoG velocities while keeping other pieces, e.g. those due to long-wavelength perturbations and small-scale density perturbations, perturbative, and make contact with effective damping models in the literature. 

Our strategy will be as follows. In Section~\ref{sec:rsd}, we review the structure of the nonlinear redshift-space mapping and connect the effect of FoGs to their characteristic functions which can be reduced to single-particle characteristic functions (SPCFs) in the large-separation and independent-draw limits. In Section~\ref{sec:WPT}, we develop the perturbation theory of these SPCFs, linking their mean-field limit to effective damping models and showing that this limit is not well-motivated due to their long-wavelength responses. Finally, we estimate the sizes of both the mean and perturbations of galaxy SPCFs in the halo model in Section~\ref{sec:halo-model}, quantifying their effect on galaxy power and bispectra for realistic galaxy samples. We close in Section~\ref{sec:conclusions} and discuss the implications of our investigations. We briefly consider several empirical damping forms in Appendix~\ref{app:stochastic-generator}.

\section{Redshift-Space Distortions (RSD) and Small-Scale Velocity Statistics}
\label{sec:rsd}

\subsection{Overview of Redshift-Space Distortions and Fingers of God (FoG)}
\label{sec:overview}

The Fourier-space galaxy overdensity in redshift space $\delta_s$ is given by 
\begin{equation}
    (2\pi)^3 \delta_D(\bk) + \delta_s(\bk) = \int d^3\bx \ e^{- i \bk \cdot (\bx + \bu_\parallel(\bx))} \ (1 + \delta_g(\bx))
\end{equation}
where $\bu_\parallel = (\hn \cdot \bu) \hn$ is the line-of-sight component of the peculiar velocity $\bu = \bv / (aH)$ in Hubble units, such that its $N$-point functions are given by
\begin{equation}
    \avg{\delta_s(\bk_1) \ldots \delta_s(\bk_N)} = (2\pi)^3 \delta_D(\bk_{1N})  \ \prod_{n=1}^{N-1} \int_{\bx_n} e^{-i\bk_n \cdot \bx_n} \ \avg{ \prod_{m=1}^N e^{-i \bk_m \cdot \bu_\parallel(\bx_m)} (1 + \delta_g(\bx_m)) }.
    \label{eqn:npoint}
\end{equation}
Translation invariance, manifesting as a delta function enforcing $\bk_{1N} = \sum_{i=1}^N \bk_i$, leads to Equation~\ref{eqn:npoint} being invariant under Galilean transformations. For example, for the power spectrum and bispectrum the exponentiated velocity factors can be re-expressed into pairwise velocity differences between two points
\begin{equation}
    P(\bk):\ e^{-i\bk \cdot (\bu_\parallel(\bx_1) - \bu_\parallel(\bx_2)) }, \quad B(\bk_1,\bk_2):\ e^{-i\bk_1 \cdot (\bu_\parallel(\bx_1) - \bu_\parallel(\bx_3)) -i\bk_2 \cdot (\bu_\parallel(\bx_2) - \bu_\parallel(\bx_3)) }
    \label{eqn:pairwise}
\end{equation}
which are unchanged by constant boosts. We will come back to these particular combinations shortly. For brevity, in the following we will use the notation $\langle \cdots \rangle'$ for the expectation value with the implied delta-function factor removed.

As described in the introduction, the size of the virial motions of galaxies implies that the perturbative evaluation of Equation~\ref{eqn:npoint} breaks down at larger scales than in real space, that is in the absence of RSD. The purpose of this paper is to investigate whether it is possible to non-perturbatively re-sum contributions from these virial velocities, so that the reach of perturbation theory can be extended to its reach in real space. Schematically, we want to perform a split
\begin{equation}
    \bu^{\rm FoG} = \bu - \bu^{\rm PT}
\end{equation}
where $\bu^{\rm PT}$ is the part of a galaxy's velocity predicted by the long-wavelength perturbative expansion, and $\bu^{\rm FoG}$ are the unresolved contributions dominated by galaxy virial motions. The split between the resolved and un-resolved parts of the velocity is not unique: the examples we present in this paper will mostly focus on the virial, intrahalo motions of galaxies but, on the other hand, many authors have noted that redshift-space counterterms larger than the naive $k_{\rm nl}$ scaling can also occur in samples sensitive only to mean halo velocities, e.g. galaxy samples with low satellite fractions (see \S\ref{sec:halo-model} for further discussion), which suggests that either even the velocities of halos are subject to small-scale dynamics with a similar scaling, or that long-wavelength velocities are less perturbative than densities at the same scales  \cite{Chen2020,Ivanov2021,Baleato2025,Chen2026}. However, in order to develop the theory of FoGs we simply require that $\bu_{\rm FoG}$ captures short-scale velocities beyond the cutoff of the long-wavelength theory, independent of their precise cause.

In the construction of the split above,  $\bu^{\rm FoG}$ is a Galilean invariant quantity, with all bulk motions already encoded in the long $\bu^{\rm PT}$. In other words, its statistics cannot depend on quantities like the velocity or gravitational potential but only on their derivatives $\partial_i \bu_j, \partial_i \partial_j \Phi$ in a spatially local way. In this split construction, we can rewrite the Fourier integrand in Equation~\ref{eqn:npoint} as
\begin{equation}
    \avg{ \avg{ \prod_{m=1}^N e^{-i \bk_m \cdot \bu^{\rm FoG}_\parallel(\bx_m)} }_{\rm FoG} \times  \prod_{m=1}^N e^{-i \bk_m \cdot \bu^{\rm PT}_\parallel(\bx_m)} (1 + \delta_s(\bx_m)) }
    \label{eqn:fog-split-npoint}
\end{equation}
where the inner expectation value $\avg{\cdots}_{\rm FoG}$ is performed over the small-scale phases of the virial velocities, \textit{conditioning} on their local environments which are averaged over in the outer expectation.

In general, the inner expectation value in Equation~\ref{eqn:fog-split-npoint} is dependent on the joint conditional probability distribution of the $N$ velocities $\bu^{\rm FoG}$ and quite complex. Indeed, this expectation value is simply the joint conditional characteristic function of the line-of-sight FoG velocity $u^{\rm FoG}_\parallel = \hn \cdot \bu^{\rm FoG}$, i.e. in terms of the cumulant generating function
\begin{equation}
    \log \avg{ \prod_{m=1}^N e^{-i \bk_m \cdot \bu^{\rm FoG}_\parallel(\bx_m)} }_{\rm FoG} = \Phi_N(\{\lambda_i\} | \{O_\alpha(\bx_i)\})
\end{equation}
where we have defined $\lambda = \hn \cdot \bk$ and $O_{\alpha}$ are a set of local operators like the overdensity $\delta$ evaluated at each position $\bx_i$. However, by construction, the FoG velocities are those arising from small-scale virial motions, i.e. they are uncorrelated for pairs of galaxies separated by more than a typical halo radius
\begin{equation}
    \text{Cov}\left[\bu^{\rm FoG}(\bx), \bu^{\rm FoG}(\bx') \right]_{\rm FoG} = 0, \quad |\bx - \bx'| \gg R_h.
\end{equation}
Indeed, a stronger statement is that, in the large-separation limit \cite{Scoccimarro2004}
\begin{equation}
    \avg{ \prod_{m=1}^N e^{-i \bk_m \cdot \bu^{\rm FoG}_\parallel(\bx_m)} }_{\rm FoG} = \prod_{m=1}^N W(\lambda_i | \{O_\alpha(\bx_i)\}), \quad |\bx_i - \bx_j| \gg R_h \quad (i\neq j). 
    \label{eqn:long_factorization}
\end{equation}
Here we have defined the \textit{single-particle} characteristic function (SPCF) $W(\lambda) = \avg{ \exp(-i\lambda u^{\rm FoG}_\parallel)}_{\rm FoG}$. Importantly, the separability above is simply a statement that small-scale dynamics de-correlates at large scales, not a statement about $\Phi_N$ at arbitrary scales.

\subsection{FoG Damping of Long-Wavelength Contributions}
\label{sec:damping1}

The decoupled structure of Equation~\ref{eqn:long_factorization} suggests a dramatic simplification of the effects of FoG's in the large-separation limit. For contributions to $N$-point functions that have their support mainly at large scales, applying the factorized form to Equation~\ref{eqn:npoint} implies that their effect amounts to dressing the long-wavelength ("$l$") predictions for galaxy densities by the SPCF $(1 + \delta_{g,s})_l \rightarrow W(\lambda) (1 + \delta_{g,s})_l$. Within this language, the effective damping models in the literature (e.g., \cite{Peacock1994,Park1994,Cole1995,HattonCole1998,Scoccimarro1999,Taruya2010,BOSS:2013uda,Sanchez2017,Eggemeier2025,Bansal2026,Euclid:2026cpu}) can be obtained by making the simplification
\begin{align}
    \Big\langle \prod_{i=1}^N W(\lambda_i | \{O_\alpha(\bx_i)\}) \ & e^{-i\bk_i\cdot \bu^{\rm PT}_\parallel(\bx_i)} (1 + \delta_g(\bx_i)) \Big\rangle_l \nonumber \\
    &\rightarrow \prod_{j=1}^N \bar{W}(\lambda_j) \times \avg{ \prod_{i=1}^N e^{-i\bk_i\cdot \bu^{\rm PT}_\parallel(\bx_i)} (1 + \delta_g(\bx_i)) }_l \quad (\text{Mean-Field ansatz})
\end{align}
where $\bar{W}$ is the mean SPCF averaging over the long-wavelength environment. Under this mean-field ansatz, the long-only contributions to galaxy $N$-point functions is simply multiplied by one power each of $\bar{W}(\lambda_i)$ for each external leg $\bk_i$. It is important to note that the mean-field ansatz does not capture all possible contributions from the SPCF even in the large-separation limit, since it neglects correlations with the large-scale environment. It is thus only strictly true in the limit that $\bu^{\rm FoG}$ is not related to the environment, e.g. in the case of large redshift errors in quasars.\footnote{However, we note that even in the case of quasars, the approximation is broken if redshift errors depend on quasar properties and thus the environment.} However, the mean-field ansatz is already very suggestive of the phenomenology of FoGs in large-scale structure; expanding in the cumulants
\begin{equation}
    \bar{W}(\lambda) = \avg{e^{-i\lambda u^{\rm FoG}_\parallel}} = \exp\left( -\frac12 (k\mu)^2 \sigma^2_{\rm FoG} + \frac{1}{24} (k\mu)^4 \kappa_{\rm FoG} + \cdots \right)
\end{equation}
In other words, the mean SPCF (or damping function), parametrized via phenomenological ansatz's like Gaussians $\exp(-\sigma^2 (k\mu)^2/2)$ or Lorentzians $(1 + \sigma^2 (k\mu)^2/2)^{-1}$, imply specific relations among the cumulants of the virial velocities of galaxies.

\subsection{Small-Scale Correlations and Cluster Expansion}
\label{sec:cluster-expansion}

The discussion above becomes more complicated for small-scale correlations at the halo scale, often called stochastic contributions. In the EFT approach, these contributions derive from short modes $\epsilon$ below the cutoff of the theory and are expanded as contact terms $\langle \epsilon \epsilon \rangle \sim \delta_D(\bx)$. However, naively applying the delta-function expansion and combining with the FoG joint characteristic function as in the large-separation case has a surprising result: looking at the pairwise expression in Equation~\ref{eqn:pairwise}, bringing two points weighted by stochastic fields $\epsilon$ results in the cancellation of the FoG velocities, leading to no damping of these stochastic contributions over the $\bk_i$ legs containing them. This apparent cancellation, however, is the result of neglecting the small-scale correlations of $\bu^{\rm FoG}$, which themselves also possess structure below $R_h$. This structure is encoded in the joint characteristic function of the velocities and its correlation with the environment, i.e. in the conditional.

In order to make further progress it is useful to make a further simplifying ansatz to the structure of the joint characteristic function. To begin, we can observe that the cancellation of the FoG factor happens exactly for a self pair, i.e. $\bx_i = \bx_j$. In order to disentangle the role of FoGs in nearby ($\lesssim R_h$) galaxies it is thus useful to briefly switch to a \textit{discrete} description of the galaxy density field, i.e.
\begin{equation}
    \delta_{g,s}(\bk) = \frac{1}{\bar{n}} \sum_i e^{-i(\bk\cdot\bx_i + \lambda u_i)}\,, \quad  \langle \delta_{g,s}(\bk_1) \cdots \delta_{g,s}(\bk_n)  \rangle = 
     \frac{1}{\nbar^n} \sum_{i_1 \cdots i_n} 
     \langle e^{-i(\bk_1\cdot\bx_{i_1} + \lambda_1 u_{i_1} + \cdots + \bk_n\cdot\bx_{i_n} + \lambda_n u_{i_n})} \rangle.
\end{equation}
where we have used the shorthand $u = \hn \cdot \bu$.

The discrete description of the galaxy field allows us to distinguish the velocity of individual galaxies situated closer than the cutoff of the long-wavelength description $R_h$. The velocities of galaxies within halos are typically not uncorrelated, but only weakly so. Roughly speaking, the relaxation time of a particle that has fallen into a halo is related to the time for a pericentric passage $\tau_{\rm relax} \propto R_h / v_{\rm vir}$ \cite{Binney2008}. This, compared to the accretion of mass onto a halo on Hubble time scales, gives that the decorrelation is roughly of order $R_h / \sigma_u \ll 1$. 

We can use this hierarchy of correlations to simplify the conditional joint characteristic function of FoG velocities. In particular, the joint characteristic function admits a cluster expansion (see e.g. \cite{Kardar2006})
\begin{equation}
    \Phi_N(\{\lambda_i\}|\{O_\alpha(\bx_i)\}
    )= \sum_i \log W(\lambda_i|\{O_\alpha(\bx_i)\}) + \sum_{i < j} \log W_2(\lambda_i, \lambda_j|\{O_\alpha(\bx_i)\}) + \cdots
    \label{eqn:cluster}
\end{equation}
where each subsequent term in the series consists of cumulants with more distinct sets of $\{\lambda_i\}$. For example, we have that
\begin{equation}
    \log W_2(\lambda_1, \lambda_2|\{O_\alpha(\bx_i)\}) = \sum_{n,m > 0} \frac{(-i)^{(n+m)} \lambda_1^n \lambda_2^m}{n! m!} \kappa_{nm}
\end{equation}
where we have defined the cumulant $\kappa_{nm} = \langle (u^{\rm fog}_1)^n (u^{\rm fog}_2)^m \rangle_c^{\rm FoG}$ In the limit that each FoG velocity is an independent draw, the series terminates at the first term, and in general subsequent terms should be suppressed by $(R_h / \sigma_u)^n$. 
In the \textit{independent draw limit}, therefore, we can dress individual galaxies with the SPCF without worrying about correlations between them.

Let us note two important caveats to the cluster expansion above. First, the characteristic functions we have written down are \textit{conditional} on each galaxy's environment. The unconditional characteristic function does not allow for the independent draw approximation, since the parameters describing their probability distributions are correlated via both long and short (intra-halo) modes. Secondly, we note that the correlated-draw terms are suppressed by $R_h/\sigma_u$, which is weaker than the naive $R_h^2 / \sigma_u^2$ separating the gradient contributions from FoG damping v.s. real-space nonlinearities---however, for the large-separation terms, this suppression is not needed and the independent draw approximation is exact.

We can now look at the power spectrum in this language. Separating into distinct and self-pairs we have
\begin{align}
    P_s(\bk) &= \frac{1}{\nbar^2} \sum_{i=j} \left\langle  e^{-i ((\bk_1+\bk_2) \cdot \bx_i + (\lambda_1 + \lambda_2) u_i))} \right \rangle' +  \frac{1}{\nbar^2}  \sum_{i\neq j} \left\langle  e^{-i (\bk_1 \cdot \bx_i + \bk_2 \cdot \bx_j + \lambda_1 u_i + \lambda_2 u_j)} \right \rangle' \nonumber \\
    &= \frac{1}{\nbar} + \frac{1}{\nbar^2} \sum_{i\neq j} \left\langle  e^{-i (\bk_1 \cdot \bx_i + \bk_2 \cdot \bx_j + \lambda_1 u^{\rm PT}_i + \lambda_2 u^{\rm PT}_j)} e^{\Phi_2(\lambda_1,\lambda_2)} \right \rangle'.
\end{align}
The self-pair term is unaffected by the FoG velocities since $\lambda_{12} = 0$ on shell, and its value is purely given by the Poissonian shot noise. The distinct pair term is what we can compute within the EFT, including shot noise beyond the Poisson regime. If we truncate Equation~\ref{eqn:cluster} in the independent-draw limit, then we have simply
\begin{equation}
    P_s(\bk) = \frac{1}{\nbar} + \langle \delta^{\rm dress}_{g,s}(\bk) | \delta^{\rm dress}_{g,s}(\bk') \rangle'
    \label{eqn:Ps_}
\end{equation}
i.e. the redshift-space density is simply dressed by the local SPCF of the FoG's
\begin{equation}
    \delta^{\rm dress}_{g,s}(\bk) = \int d^3 \bx \ e^{-i\bk \cdot \bx - i \lambda \bu^{\rm PT}(\bx)} W(\lambda | \bx) (1 + \delta_g(\bx)).
    \label{eqn:delta-dress}
\end{equation}
Under the mean-field ansatz, this corresponds to damping all but the Poisson shot noise contribution to the galaxy power spectrum, i.e. we see that the $\nbar^{-1}$ plays a privileged role in this expansion---we note, however, that the mean-field ansatz does not include, among other effects, the correlated short-scale noise in $W(\lambda|\bx)$, that can change this simple picture of FoGs and the stochastic power spectrum. Of course, we can also systematically include higher particle-number correlators in the cluster expansion by including the operator $e^{\Phi_2}$ in order to account for nontrivial correlations at short scales.

We can similarly evaluate the bispectrum and higher-order $N$-point functions. Schematically,
\begin{align}
    &\bar{n}^3 B_s(\bk_1, \bk_2) \nonumber \\
    &= \bar{n}^3 \langle \sum_{i=j=k} \cdots \rangle + \bar{n}^3 \big( \langle \sum_{i=j\neq k} \cdots \rangle + \text{et cycl} \big) + \bar{n}^3 \langle \sum_{i \neq j \neq k} \cdots \rangle \nonumber \\
    &= \nbar + \nbar^2 \left \langle \sum_{i=j\neq k} e^{-i(\bk_1 + \bk_2)\cdot \bx_i -i (\lambda_1 + \lambda_2) u_i } e^{-i\bk_3 \cdot \bx_k -i \lambda_3 u_k } + \text{et cycl} \right \rangle + \nbar^3  \langle \sum_{i \neq j \neq k} \cdots \rangle \nonumber \\
    &= \nbar + \nbar^2 \left( P^{\rm red}_s(\bk_1) + P^{\rm red}_s(\bk_2) + P^{\rm red}_s(\bk_3) \right) + \nbar^3 \langle \sum_{i \neq j \neq k} \cdots \rangle
\end{align}
where in the last line we have used that $\bk_1 + \bk_2 = -\bk_3$ and we have defined $P_s^{\rm red} = P_s - \nbar^{-1}$ to be the power spectrum with self pairs removed. In most practical analyses the first two terms, which have all and one powers of the FoG damping removed because all or one pair of indices are coincident, are removed exactly based on the estimated shot noise and power spectrum. In the independent draw limit, the remaining bispectrum is simply then the bispectrum of the dressed field
\begin{equation}
    B_s^{\rm red}(\bk_1, \bk_2) = \langle  \delta^{\rm dress}_{g,s}(\bk_1) \delta^{\rm dress}_{g,s}(\bk_2) \delta^{\rm dress}_{g,s}(\bk_3) \rangle'.
    \label{eqn:Bs_}
\end{equation}
In the above expression we have used the superscript ``red'' to refer to the bispectrum with the shot-noise terms removed---we will omit these terms and distinction in the rest of the text unless otherwise noted.

\section{Perturbation Theory of the Characteristic Function}
\label{sec:WPT}

We want to now explicitly construct the SPCF $W(\lambda|\bx)$ in the effective field theory. Since $W$ characterizes the varying distribution of small-scale velocities at each point, it is not a simple function of $\lambda$ but rather a functional also of the local environment. The goal of this section is to formalize this in terms of responses to the large-scale structure of the universe, i.e. in terms of operators $O_{L}(\bx)$ constructed from long-wavelength modes, in addition to stochastic contributions from short-wavelength fluctuations.

The long-wavelength responses of the SPCF follow a similar logic to the widely used bias expansion of galaxies, where e.g. the response to a long-wavelength density perturbation $\delta_{L}$ is controlled by the linear bias $b_1$ (see e.g. ref.~\cite{bias-review}). A notable difference, however, is that, since $W(\lambda)$ is constructed from the line-of-sight projection of the peculiar velocity, its responses do not have to be limited to isotropic, scalar operators like the density $\delta_L$. More explicitly, starting with the one-velocity term, we have
\begin{align}
    W(\lambda) &= 1 - i\lambda \avg{u}_{\rm fog} - \frac12 \lambda^2 \avg{u^2}_{\rm fog} + \ldots \nonumber \\
    &= 1 - i \hat{N}_{ij} \bk_i \avg{\bu^{\rm FoG}_j}_{\rm fog} - \frac12 \hat{N}_{ijkl} \bk_i \bk_j \avg{\bu^{\rm FoG}_k \bu^{\rm FoG}_l}_{\rm fog} + \cdots
\end{align}
where we have defined the tensors $\hat{N}_{ijk...} = \hn_i \hn_j \hn_k \cdots$. The products of velocities are tensor fields, and their long-wavelength responses therefore take the form
\begin{equation}
    \hn_i \hn_j \langle \bu^{\rm FoG}_i \bu^{\rm FoG}_j \rangle_{\rm fog} \supset \hn_i \hn_j O_{L,ij}(\bx) \equiv O_{L,\parallel}(\bx).
\end{equation}
Note that the tensor structure of $O_{L,ij}$ need not be complicated, e.g. $O_{L,ij} = \delta_{L} \delta_{ij}$ has $O_{L,\parallel} = \delta_L$.

We can now write down the effect of these responses on the linear-theory prediction for the redshift-space galaxy density field. To lowest order, we can write the SPCF including the linear density $\delta$ and line-of-sight tidal fields $s_\parallel = \hn_i \hn_j s_{ij}$. The latter can also be recast as the line-of-sight velocity gradient $\eta = f (s_\parallel + \delta/3)$, where $f$ is the linear growth rate. In other words, we have
\begin{equation}
    W(\lambda) = \bar{W}(\lambda) \left( 1 + R_\delta(\lambda)  \delta + R_\eta(\lambda)  \eta + \ldots \right) = \bar{W}(\lambda)  + W_\delta(\lambda)  \delta + W_\eta(\lambda)  \eta + \ldots.
\end{equation}
where we have defined absolute and relative responses to each operator $O$ as $W_O = \bar{W} R_O$. These responses modify the FoG-resummed linear redshift-space density field to
\begin{equation}
    \delta^{\rm lin, res}_{g,s}(\bk) = \bar{W}(\lambda) \left( b_1 + R_\delta(\lambda)  + f \left(1 + R_\eta(\lambda) \right) \mu^2  \right)\ \delta_0(\bk).
    \label{eqn:res-linear-theory}
\end{equation}
In other words, in order to properly resum the effects of FoG, it is insufficient to simply account for the damping due to the mean SPCF $\bar{W}$; rather, the small-scale velocities responsible for FoG have long-wavelength responses that induce different scale dependences for different terms in the perturbative expansion, in this case the isotropic and anisotropic pieces of the linear-theory prediction. This phenomenon was first noticed by ref.~\cite{Seljak2001} in the context of the halo model---we will revisit these halo-model calculations in the EFT context in Section~\ref{sec:halo-model} and show that these responses generically produce order-one corrections to the mean SPCF prediction, and that the deviations worsen for higher-order bias. In configuration space, the fact that the response coefficients are not constants but functions of $k\mu$ amount to the convolution of the corresponding bias operators with a line-of-sight kernel which satisfies particular rules we will derive next. Equivalently, at linear order
\begin{equation}
 Z_1^{\rm FoG}(\bk)=\bar{W}(k_\parallel)Z_1^{\rm K}(\bk) +W_\delta(k_\parallel) +f W_\eta(k_\parallel) \mu^2  
\end{equation} where $Z_1^{\rm K}(\bk)=b_1+f\mu^2$ is the linear Kaiser kernel. This makes explicit the separation between the mean FoG contribution and its long-wavelength response terms.

\subsection{Symmetries and Selection Rules}

At first glance, Equation~\ref{eqn:res-linear-theory} might suggest a modification of the Kaiser formula via long-wavelength FoG responses. In fact, the modification on large scales $k \sigma_u \ll 1$ is forbidden by symmetry constraints on the SPCF $W(\lambda|\{O_L^i(\bx)\})$, which require $\bar{W} \rightarrow 1$, $R_{\delta,s} \rightarrow 0$ in this limit. In fact, the leading scale-dependence of the responses $R_O$ at low $\lambda$ is strongly constrained by the tensorial properties of each operator $O$.

More specifically, we can use the fact that $W$ obeys the following symmetries and constraints
\begin{enumerate}
    \item \textit{Parity}. The SPCF is invariant under the transformation $\bx \rightarrow -\bx$, $\bu \rightarrow -\bu$. Note that, since $W(\lambda)$ always multiplies a mode with wavevector $\bk$, we also need to explicitly enforce $\bk \rightarrow -\bk, \lambda \rightarrow -\lambda$ in the transformation.
    \item \textit{Line-of-sight reversal}. Since $\hn$ enters twice in the redshift-space mapping, the redshift-space field is invariant under $\hn \rightarrow -\hn$, which also takes $\mu, \lambda \rightarrow -\mu, -\lambda$. 
    \item \textit{Conservation of probability}. The value of $W(0) = \langle e^{-i \lambda u} \rangle_{\lambda=0} = 1$.
    \item \textit{Rotations}. Before fixing $\hat{n}$, the theory is $SO(3)$ covariant. Fixing the line of sight selects a preferred direction,  leaving only rotations about $\hat{n}$.
\end{enumerate}
Since $\bu$ is real, we also have $W(\lambda)^\ast = W(-\lambda)$. Translated into constraints on the response to a long-wavelength operator $O(\bx)$, these constraints imply e.g. $W_O(\lambda)^\ast = (-1)^{P_O} W_O(\lambda)$, $W_O(-\lambda) = (-1)^{P_O} W_O(\lambda)$, and $W_O(\lambda)^\ast = W_O(-\lambda)$, where $P_O$ is the parity of the operator $O(\bx)$. Importantly, parity-odd operators are not immediately forbidden in this expansion due to their coupling with the wavevector, though we will see that their contributions are suppressed from effective-theory arguments. These constraints are the line-of-sight projected equivalent of the general selection rules for redshift-space tensor fields, e.g. galaxy shapes in spectroscopic surveys, recently derived in ref.~\cite{Akitsu26}.

Indeed, the above properties become manifest when we consider the explicit construction of the SPCF $W$
\begin{equation}
    W(\lambda|\{O_i(\bx)\}) \supset (-i \lambda)^N \hn_{i_1} \cdots \hn_{i_N} \langle u^{\rm FoG}_{i_1} \cdots u^{\rm FoG}_{i_N} \rangle_{\rm fog} \supset (-i \lambda)^N O_i^{M\leq N}(\bx)
\end{equation}
where $M$ is the number of insertions of $\hn$ into the operator $O_i$, less than or equal to $N$ and equal to it modulo $2$ since accounting for pairs that contract with $\delta_{ij}$'s to unity. This implies that we can write
\begin{equation}
    W_O(\lambda) = (-i \lambda)^M \hat{W}_O(\lambda^2)
    \label{eqn:los-counting}
\end{equation}
where $\hat{W}_O(\lambda^2)$ is analytic at $\lambda^2 = 0$. This structure, imposed purely by the redshift-space mapping itself, is sufficient to recover all of the above constraints on $W$ and $W_O$. Considering probability conservation, $W_O$ starts at order $\lambda^2$ for scalar responses ($M=0$), while for tensor responses with $M\geq 1$ it starts at order $\lambda^M$.

\subsection{Bias Expansion of the SPCF}

The virial velocities of galaxies, and hence the SPCF $W(\lambda|\bx)$ can depend only locally in space on the environment of the galaxy, though they are also naturally sensitive to the formation history of halos, and thus can depend on the environment non-locally in time. In addition, the environmental dependence must satisfy the Equivalence Principle and Galilean invariance, i.e. cannot depend on the bare gravitational, its first derivative, or bare velocities. A general bias expansion of line-of-sight dependent galaxy density fields was formulated in ref.~\cite{Desjacques2018} based on the operators
\begin{equation}
    \Pi^{[n]}_{ij} = \frac{1}{(n-1)!} \left( \frac{1}{\mathcal{H} f} \frac{D}{D\tau}  \Pi^{[n-1]}_{ij} - (n-1)  \Pi^{[n-1]}_{ij} \right)
\end{equation}
where $\Pi^{[1]} = s_{ij} + (\delta_{ij}/3) \delta  \propto \partial_i \partial_j \Phi$ is the second derivative of the gravitational potential and $D/D\tau = \partial_\tau + v_i \partial_i$ is the convective derivative. Each $\Pi^{(n)}$ is order $n$ at leading order, and together they constitute the full set of local in space but nonlocal in time operators long-wavelength responses can depend on. In order to arrive at the anisotropic basis, one simply forms the set of all possible contractions with the Kronecker-$\delta$, as would be the case for the general isotropic bias basis, and powers of the line of sight $\hn_i$. For example, up to second order in the initial conditions this is equivalent to the combinations
\begin{enumerate}
    \item 1$^{st}$ order. $\delta = \text{Tr} \Pi^{[1]}$, $\eta = f \Pi^{[1]}_\parallel$
    \item 2$^{nd}$ order. $\delta^2$, $s^2$, $\delta \eta$, $\eta^2$, $(ss)_\parallel$, $\Pi^{[2]}_\parallel = \hn_i \hn_j \left( \left( \Pi^{[1]} \right)^2_{ij} + \frac{10}{21} \frac{\partial_i \partial_j}{\partial^2}(\delta^2 - \frac32 s^2) \right)$
\end{enumerate}
In Table~\ref{tab:fog-response-basis}, we list all possible operators up to cubic order as in ref.~\cite{Desjacques2018}, noting moreover the suppression of the coefficients in $\lambda$ by selection rules. We refer the interested reader to the detailed definition of each operator, and the evaluation of their statistics at the level of loops, to that work. It is important to note that the bias expansion above is formulated in real space, i.e. in order to make contact with observations each operator in addition has to be mapped to its observed redshift-space coordinate due to $\bu^{\rm PT}$. In addition, at 1-loop order, many of the cubic operators enter in degenerate ways; in the case of the SPCF, this is complicated by the selection rules in $\lambda$, such that otherwise degenerate operators may enter at different powers of $\lambda$ at leading order. In this case, the leading scaling of the degenerate combination would be governed by the contributing operator with lowest $M$. We leave the full enumeration of loop contributions to future work.

\begin{table}[t]
\centering
\renewcommand{\arraystretch}{1.35}
\begin{tabular}{c|c|l|c}
\hline
$\,\quad n \quad\, $ & $\quad  M \quad $ & Operators $O$ & Leading $W_O(\lambda)$ \\
\hline\hline
$1$ & $0$ & $\delta$ & $\lambda^2$ \\
$1$ & $2$ & $\eta$ & $\lambda^2$ \\
\hline
$2$ & $0$ & $\delta^2,\ s^2$ & $\lambda^2$ \\
$2$ & $2$ & $\delta\eta,\ (ss)_\parallel,\ \Pi^{[2]}_\parallel$ & $\lambda^2$ \\
$2$ & $4$ & $\eta^2$ & $\lambda^4$ \\
\hline
$3$ & $0$ & $\delta^3,\ \delta s^2,\ s^3,\ O_{\rm td}$ & $\lambda^2$ \\
$3$ & $2$ & $s^3_\parallel, \, \eta\delta^2,\ \eta s^2,\ \delta(ss)_\parallel,\ \delta\Pi^{[2]}_\parallel,\ (s\Pi^{[2]})_\parallel,\ \Pi^{[3]}_\parallel$ $\qquad$ & $\lambda^2$ \\
$3$ & $4$ & $\eta^2\delta,\ \eta(ss)_\parallel,\ \eta\Pi^{[2]}_\parallel$ & $\lambda^4$ \\
$3$ & $6$ & $\eta^3$ & $\lambda^6$ \\
\hline
\end{tabular}
\caption{Dimensionless deterministic response operators through third order, organized by perturbative order $n$ and line-of-sight degree $M$. For $M\geq1$, $W_O(\lambda)=(-i\lambda)^M\hat W_O(\lambda^2)$, while normalization of the conditional SPCF implies that nontrivial scalar responses with $M=0$ begin at $\lambda^2$.}
\label{tab:fog-response-basis}
\end{table}

Let us comment briefly on the connection of the long-wavelength expansion of the SPCF to other observables in large-scale structure. Within the context of the EFT of galaxy bias line-of-sight dependent terms can arise on large scales due to observer-dependent selection effects \cite{Hirata2009}, which would correspond in our case to the forbidden $R_\eta \rightarrow \text{const.}$ in Equation~\ref{eqn:res-linear-theory}. However, since the long-wavelength responses $R_O$ are in general functions of $\lambda=k\mu$, satisfying Equation~\ref{eqn:los-counting}, their appearance is not forbidden, as long as they are always accompanied by minimum powers of $k\mu$. For example, the operator $\eta$ has two insertions of $\hn$, and therefore must appear at order $(k\mu)^2$ and, similarly, since $W=1$ for $\lambda = 0$, even the scalar density response $R_{\delta}$ must appear at order $(k\mu)^2$, though we note that in the latter case the \textit{total} $b_1 + R_O$ does not have this limitation. It is also worth noting that these selection effects arise also in the Lyman-$\alpha$ forest, where they are similarly generated by the optical-depth to flux exponential $F = e^{-\tau}$; since $\tau$ is a redshift-space, field, the exponential mapping naturally generates anisotropic responses \cite{Seljak2012,Chen2021,Ivanov2024}. In contrast to the FoG SPCF, no powers of the wavenumber $k$ multiply $\tau$, leading to response coefficients that can approach constants at large scales. This is a manifestation of number conservation in the FoG case.

\subsection{Counterterms and Stochastic Parameters}
\label{ssec:cts}

The anisotropic bias basis above encompasses the dimensionless bias operators in the expansion. Beyond these, it is also necessary to include higher derivative operators, or counterterms. Indeed, unlike the typical galaxy bias expansion, these terms can begin at the \textit{first} derivative, i.e. 
\begin{equation}
    W(\lambda | \bx) \supset \bar{W}(\lambda) R_{\partial_\parallel \delta}(\lambda) \partial_\parallel \delta_L = (-i\lambda) \bar{W}(\lambda) \hat{R}_{\partial_\parallel \delta}(\lambda^2) \partial_\parallel \delta_L.
\end{equation}
This is because only $W$ needs to be invariant under parity, not the individual operators it is composed of, given that $\lambda$ also has odd parity. Physically, this is equivalent to the statement that the virial velocities of halos in a large-scale gradient can be asymmetric. However, at 1-loop order in the power spectrum, $(-ik\mu) \partial_\parallel \delta = (k\mu)^2 \delta$, so that its contribution is degenerate with the even operators
\begin{equation}
    W(\lambda | \bx) \supset \bar{W}(\lambda) \left( R_{\partial^2 \delta} \partial^2 \delta + R_{\partial^2 \eta} \partial^2 \eta + R_{\partial_\parallel^2 \eta} \partial_\parallel^2 \eta \right)
\end{equation}
i.e. the usual 1-loop counterterms in the redshift-space galaxy power spectrum, except with scalar coefficients promoted to functions of $\lambda$. Indeed, the purely gradient terms encoding the perturbative spatial nonlocality at the halo scale should be suppressed by $R_h$ compared to terms that are purely proportional to the virial scale $\sigma_u^2$, similar to the degenerate contribution of stochastic velocities to the first and second pairwise velocity spectra \cite{Chen2020}. Moreover, note that the leading linear-in-$\lambda$ contribution to $R_{\partial_\parallel \delta}$ constitutes a nonzero conditional \textit{mean}, i.e. $\avg{u^{\rm FoG}_\parallel} \supset \text{const.} \times \partial_\parallel \delta$. Such a nonzero mean reflects an inherent ambiguity in the split between $\bu^{\rm FoG, PT}$; were this deterministic mean relegated to $\bu^{\rm PT}$, e.g. because it is perturbative compared to other FoG terms as we have argued, then $R_{\partial_\parallel \delta}$ would instead begin at $\lambda^3$, i.e. at the level of a skewness in halo velocities generated by a local density gradient.

Finally, in addition to the deterministic responses we have written down so far, much like the galaxy density field itself the finger-of-god characteristic functions should also carry stochastic fluctuations. For example, halos that form under equal long-wavelength environments may nonetheless have different velocity dispersions due to small-wavelength fluctuations at the halo scale. At the level of the power spectrum, it is sufficient to write
\begin{equation}
    W(\lambda|\bx) = \bar{W} \left(1 + \delta W_{\rm det} + \epsilon_W \right)
\end{equation}
such that the contribution to the galaxy power spectrum is
\begin{equation}
    P_{g,s}(\bk) \supset \bar{W}^2(\lambda) \langle \epsilon_g(\bk) | (\epsilon_g \epsilon_W)(\bk') \rangle'.
\end{equation}
The leading $(k\mu)^2$ term above expresses the excess velocity dispersion of the FoG velocities conditioning on nearby pairs encoded by the galaxy density-field stochasticity $\epsilon_g$. This correlation is not naturally suppressed--- since it inhabits a volume $R_h^3$ and has characteristic size $\sigma_u^2$, it has in principle a similar scaling to the FoG-induced scale-dependence $P_{\rm shot} \sigma_u^2 (k\mu)^2 / \nbar $. We are therefore left with a free function
\begin{equation}
    P_{\rm stoch}(\bk) = \frac{1}{\nbar} \left(1 + P_{\rm shot} + a_2 k^2 + \sum_{n=1}^\infty (a^{\rm fog}_{0n} + a^{\rm fog}_{2n} k^2) (k\mu)^{2n} \right)
\end{equation}
where the damping of the non-Poisson pieces is absorbed into the power series in $k\mu$.
This is similar to the situation for the deterministic responses---in this case, the operator shape is simply constant, but we still gain a free function of $\lambda$ whose scale-dependence is governed by $\sigma_u^2$. We caution, however, that this does not amount to adding an arbitrary function to the power spectrum, since the free function is purely a function of $k_\parallel = k\mu$. The case of the long-short contributions in the tree-level bispectrum is analogous. While we defer careful characterizations of the stochastic power spectrum and bispectrum of FoG velocities to future work, we will revisit the behavior of these stochastic terms in the context of halos in Section~\ref{sec:halo-model}.

\subsection{FoGs in Practice}

We close this section with some practical observations about FoG modeling. In recent years, EFT models have become the state-of-the-art for modeling redshift-space galaxy clustering. Our explorations so far on resumming the SPCF builds on this progress: perturbations in $W(\lambda|\bx)$ are expanded, and $N$-point functions are computed to a given loop order including these perturbations, if at the cost of introducing new bias operators and free functions.

However, it is important to note that, at a given loop order, some subset FoG effects are \textit{already} accounted for by EFT counterterms, since the EFT is an asymptotically complete theory. For example, at 1-loop in the power spectrum we have $\delta_{\rm ct} \supset c_2 k^2 \mu^2 \delta + c_4 k^2 \mu^2 \eta$ which is degenerate with the linear responses of the SPCF to $\delta$ and $\eta$. In the limit that all enhanced nonlinearities are properly encoded in the resummation, the natural priors on the former are of order the nonlinear scale $k_{\rm nl}^{-2}$, while the latter are of order $\sigma_u^2$,  allowing the two sets to be distinguished by their natural power counting. In a fit adopting such priors, $c_{2,4}$ will be limited in size and unable to account for the empirically observed damping of the power spectrum by FoG.  However, when FoGs are described in the mean field, only one $\lambda^2$ degree of freedom is available in the mean SPCF, i.e. from the variance of the FoG distribution, meaning that at least one of $c_{2,4}$ needs to be set at order $\sigma_u^2$ to compensate at $\mathcal{O}(\lambda^2)$.

In the limit that both $c_{2,4}$ are given priors of order $\sigma_u^2$, they can fully compensate for the leading $\mathcal{O}(\lambda^2)$ behavior of the linear SPCF responses, meaning that any non-degenerate FoG contributions arise at $\mathcal{O}(\lambda^4)$. Since there are two linear responses $\propto \lambda^4, \lambda^4 \mu^2$ at this next order, this generally requires two additional degrees of freedom on top of the $\mathcal{O}(\lambda^2)$ counterterms, implying that a 1-parameter effective damping (e.g. the exponential, Lorentzian or VDG forms) is generally insufficient at this order, while a 2-parameter model would make up the difference.\footnote{However, we note that by matching the $\lambda^4$ responses rather than the $\lambda^4 \mu^2$ one, the difference in the 1-parameter case can be sequestered to terms scaling as $\mu^6$ at this order in $\lambda$.}

Two further notes are in order. First, the above discussion about the degeneracy between 1-loop counterterms neglects the contribution of nonlinear SPCF responses, which is technically not consistent in the counting where $\lambda \sigma_u$ nonlinearities are enhanced, e.g. terms like $\propto \lambda^2 P_{\delta^2 \delta^2}$ are as large as 1-loop terms even though they are 2-loop in the naive EFT power counting, since they are enhanced by  $(k_\text{nl} \sigma_u)^2$. Second, the ability of counterterms to soak up SPCF responses in the mean-field ansatz requires that they are also damped by $\bar{W}$. Were they not to be included in the FoG damping, the only possible contribution would be of the form
\begin{equation}
    \delta_{\rm fog} \supset c_{\rm fog} (k\mu)^4 (b + f\mu^2)
    \label{eqn:cfog}
\end{equation}
independent of the number of parameters describing the damping function, since the only possible $\lambda^4$ contribution would be proportional to the Kaiser kernel. This suggests that it is not sufficient in general to add only FoG contributions damping the linear term. Indeed, it can be readily seen that the above argument holds even when substituting $c_{\rm fog}$ for any generically parametrized $\bar{W}$, which also includes as a subset the measure-zero subspace where $c_{2,4}$ are exactly proportional to the Kaiser kernel.

Finally, let us comment on an extension of effective damping models with small numbers of free parameters that nonetheless captures the long-wavelength responses of the SPCF. This can be achieved by promoting effective damping parameters, e.g. the width of a Lorentzian, to local rather than global parameters, such that their long-range correlations can be described by a bias expansion. An attractive feature of parametrizing the SPCF with small numbers of free parameters is that the long-wavelength responses of the SPCF then collapse into a finite set of templates at a given perturbative order, rather than an infinite set of functions. For example, in the 1-parameter case, which describes most models in the literature, any environmental dependence of the SPCF must be governed by
\begin{equation}
    W(\lambda|\bx) = \bar{W}_{\theta(\bx)}(\lambda) , \quad \frac{\partial W(\lambda|\bx)}{\partial O(\bx)} = \left( \frac{\partial \theta(\bx)}{\partial O(\bx)} \right)\  \partial_\theta \bar{W} \equiv b^\theta_O\ \partial_\theta \bar{W}(\lambda).
\end{equation}
In this case we do not gain a series of unknown response functions but merely a set of unknown response coefficients $b^\theta_O$ of the environmental variable $\theta(\bx)$, multiplying a functional template $\partial_\theta \bar{W}(\lambda)$. A higher-dimensional parametrization of $W$, for example one that depends on both the mean velocity dispersion and satellite fraction (\S~\ref{sec:halo-model}) simply results in more templates, and the generalization to higher order bias is straightforward. However, by necessity, this restriction implies that the responses cannot be fit at all orders in $\lambda$, making it more attractive when $\lambda \sigma_u \lesssim 1$.

\section{Estimates from the Halo Model}
\label{sec:halo-model}

In Section~\ref{sec:WPT}, we developed a perturbation theory for the single-particle characteristic function of FoG velocities. Taken at face value, this theory shows that non-perturbatively resumming the effects of FoGs on galaxy clustering requires, in principle, including the anisotropic responses of the FoG velocity distributions to the background environment, leading to a significant expansion of the set of bias operators necessary along with coefficients dependent on $\lambda = k\mu$. However, the question remains whether these responses are significant for realistic galaxy samples, i.e. those observed by spectroscopic surveys like DESI. In this section, we answer this question in the affirmative by estimating the responses within the halo model using the peak-background split \cite{Sheth1999}.

\subsection{Formalism}

A sample consisting of distinct populations $\alpha$ has a SPCF given by the weighted sum
\begin{equation}
    W(\lambda) = \sum_{\alpha} f_\alpha \avg{e^{-i\lambda u^{\rm fog}_\alpha}}_{\rm FoG}.
\end{equation}
For example, in the halo model where populations are distinguished by mass and into centrals and satellites, we have
\begin{align}
    W_{\rm HOD}(\lambda) &= \frac{1}{\bar{n}_g} \int dM\ n(M)\ \left( N_c(M) W_{c|M}(\lambda ) + N_s(M) W_{s|M}(\lambda) \right) \nonumber \\
    &= \frac{1}{\bar{n}_g} \int dM\ n(M) N_g(M) W_M(\lambda)
\end{align}
where $W_{\alpha |M}$ is the SPCF for each individual population and $\bar{n}_g$ is the total number density of galaxies in the sample. In the second line, $N_g(M)$ is the number of galaxies per halo per mass, and $W_M$ is the SPCF conditioning on mass. 

The HOD form of $W(\lambda)$ allows us to compute its long-wavelength responses using the peak-background split (PBS) \cite{Mo:1995cs,Sheth1999,Schmidt:2012ys}.In the PBS, the changes in galaxy properties and statistics in the presence of a long-wavelength mode are computed via their dependence on an equivalent change in the background matter density. In the simplest HOD models, the linear Lagrangian bias of a halo of mass $M$ can be expressed as a derivative of the mass function $b_1^L(M) = \partial\ln n(M)/\partial\delta_L$, such that the linear bias of the galaxy sample is 
\begin{equation}
    b_1^L = \frac{1}{\bar{n}_g} \int dM \ \left( \frac{\partial\ln n(M)}{\partial\delta_L} \right) \ n(M) N_g(M) \equiv \avg{b_1^L(M)}
\end{equation}
where in the final step the average is taken over all galaxies by integrating over the mass function and halo occupation $N_g$. In this formalism, we can equally write the linear density response of the SPCF as
\begin{equation}
    R_\delta(\lambda) = \left( \frac{ \avg{b_1^L(M) W(M)} }{\avg{W(M)}} - \avg{b_1^L(M)} \right) + \frac{\avg{\partial_\delta W(M)}}{\avg{W(M)}}.
    \label{eqn:pbs-Rdelta}
\end{equation}
Here, the first contribution comes from changing the weighting of galaxies contributing their FoGs towards heavier halos in an overdensity, while the second contribution comes from the changing of the velocity dispersion in a given mass halo. The responeses to higher-order biases like $\delta^2$ can be similarly obtained by taking further derivatives $\partial_{\delta_L}$ of the mass function and $W(\lambda)$.

Before moving on to the more realistic case, let us explore the consequences of the calculations above in a toy model HOD, where galaxies in halos of mass $M$ have Gaussian velocity distribution with dispersion given by Equation~\ref{eqn:sigmaM}. We first examine the low-$k$ behavior of $W(\lambda)$ in this toy model, where to leading order
\begin{equation}
    W_M(\lambda) = \exp\left[ -\frac12 \lambda^2 \sigma_uv^2(M) \right] = 1 -\frac12 \lambda^2 \sigma_u^2(M) + \ldots.
\end{equation}
In this limit we have that
\begin{align}
    \frac{ \avg{b_1^L(M) W(M)} }{\avg{W(M)}} - \avg{b_1^L} &= \frac{ \avg{b_1^L} - \frac12 \lambda^2 \avg{b_1^L \sigma^2_v} + \cdots}{1 - \frac12 \lambda^2 \avg{\sigma^2_v} + \cdots} - \avg{b_1^L} \nonumber \\
    &=  -\frac12 \lambda^2  \left( \avg{b_1^L \sigma^2_v} - \avg{b_1^L} \avg{\sigma^2_v} \right) + \cdots
\end{align}
i.e. the number-weighting response is proportional to the covariance of the Lagrangian bias and velocity dispersion. This covariance is positive given that the bias and velocity dispersion both scale positively with mass. 

On the other hand, the second term in Equation~\ref{eqn:pbs-Rdelta} gives \cite{Cooray2002}
\begin{equation}
    \frac{\avg{\partial_\delta W(M)}}{\avg{W(M)}} = -\frac12 \lambda^2 \avg{ \partial_\delta \sigma_u^2} + \cdots = - \frac{\alpha_v}{2} \lambda^2 \avg{\sigma_u^2} + \cdots, \quad \alpha_v = \frac{1}{3},
\end{equation}
where we have used that at fixed mass $V^2_{200} \propto R_{200}^{-1} \propto (M/\rho)^{-1/3} \propto (1 + \delta_L)^{1/3}$, since the background density is altered by $(1 + \delta_L)$.\footnote{Note that here we have defined the halo as an overdensity relative to the \textit{background} density in order to better match the setup of the DESI HODs below.} Note that both of these contributions are negative---overdensities source both more massive, and hotter (smaller) halos. When this SPCF is multiplied with the halo density field we get the effective linear bias
\begin{equation}
    b_1^{\rm eff} = \bar{W} \left( 1 + b_1^L + R_\delta \right).
\end{equation}

In general the leading-order estimates above hold only at large radial scales. For smaller $\lambda$, while $\lambda \sigma_u$ continues to grow, $W(\lambda)$ is limited to be less than one. This limits the effects of the most massive halos, for which $W(M)$ is restricted to be near 0. In practice, the exact averages over $M$ can be computed for the exponential $W_M$ by numerical integration, from which we can extract the leading coefficients $R_{\delta, \delta^2}$. Combining with the bias expansion we see that the galaxy density field in redshift space is modified to
\begin{align}
    \left [ W(\lambda) (1 + \delta_{g,s}) \right]^L &= \bar{W} \left(1 + (b_1^L + R_{\delta}) \delta + \frac12 ( b_2^L + R_{\delta^2} + 2 b_1^L R_{\delta} ) \delta^2 + (b^L_{s^2} + R_{s^2}) s^2 \right) \nonumber \\
    &= 1 + b_1^{L, \rm eff}(\lambda) \delta + \frac12 b_2^{L,\rm  eff}(\lambda) \delta^2 + b_{s^2}^{L, \rm eff}(\lambda) s^2
\end{align}
where in the second line we have defined the effective biases $b_O^{\rm eff}$ as the coefficient of each operator once FoG effects are resummed including the scale-dependent response terms. The above expression is written in Lagrangian space (``$L$'') since that is naturally where the PBS occurs; to convert to the standard Eulerian biases we can use the relations $b^E_1 = 1 + b^L_1$, $b^E_2 = b^L_2 + \frac{8}{21} b_1^L$, and $b^E_{s^2} = b^L_{s^2} - \frac{2}{7} b_1^L$ \cite{Abidi18}. More specifically, in order to obtain the FoG-resummed Eulerian bias coefficients we use the above bias translation with the scale-dependent effective bias coefficients.

Finally, in the above we have used the PBS to compute the isotropic responses of the SPCF to long-wavelength perturbations. As we showed in Section~\ref{sec:WPT}, at any given order the SPCF also admits responses to anisotropic perturbations, e.g. to background tidal fields. The estimation of these latter responses are nontrivial and will be most easily accomplished via numerical simulations. However, if we assume that halo density profiles, and thus shapes, respond in a similar fashion to halo velocity distributions, as is the case for halos in dynamical equilibrium, then it is likely that the anisotropic responses are suppressed similar to that of halo intrinsic alignments, which have a typical size of at most $\mathcal{O}(10\%)$ in contrast to halo densities which have order-one responses to long-wavelength perturbations \cite{Akitsu2023}. In fact, the bulk of the isotropic responses estimated above come from the responses of the halo mass function, which as an isotropic scalar cannot couple with anisotropic operators.

\subsection{Application to DESI HODs}

We can now apply the calculations above to explore the effect of FoGs on DESI-like galaxies. In particular, we will explore these effects in the context of the HODs for DESI luminous red galaxies (LRGs), quasars (QSOs) \cite{Yuan2024} and emission-line galaxies (ELGs) \cite{Rocher2023}. In order to compute the effects of FoGs, we will use the Sheth-Tormen mass function \cite{Sheth1999} and derive the Lagrangian linear and quadratic galaxy biases $b^L_{1,2}$ using the PBS. We will also make use of the Lagrangian tidal bias $b^L_{s^2}(M)$ as measured in ref.~\cite{Abidi18}.\footnote{Ref.~\cite{Abidi18} report the tidal bias as a function of halo mass $M$ at $z=0$; we approximate its value at higher redshifts by assuming that the value of the bias depends only on peak height $\nu$ and calibrating to the implied $b^L_1$-$b^L_{s^2}$ relation.}

Reference \cite{Yuan2024} models LRGs using a variant of the Zheng et al \cite{Zheng2005} parametrization where, roughly speaking, central galaxies occupy halos above a minimum mass $M_{\rm cut}$, and halos with centrals acquire satellites with number scaling as a power law in mass with index $\alpha$. Notably, the velocities of centrals and satellites are seeded with fractions $\alpha_\text{c,s}$ of (a) a normal distribution with the measured velocity dispersion of the halo and (b) the relative velocity between a particle and the halo center; in order to proceed analytically, we will approximate both of these distributions via a normal distribution with width $\sigma_u(M)$, with the corresponding long-mode response to $\delta_L$.

\begin{figure}
    \centering
    \includegraphics[width=\linewidth]{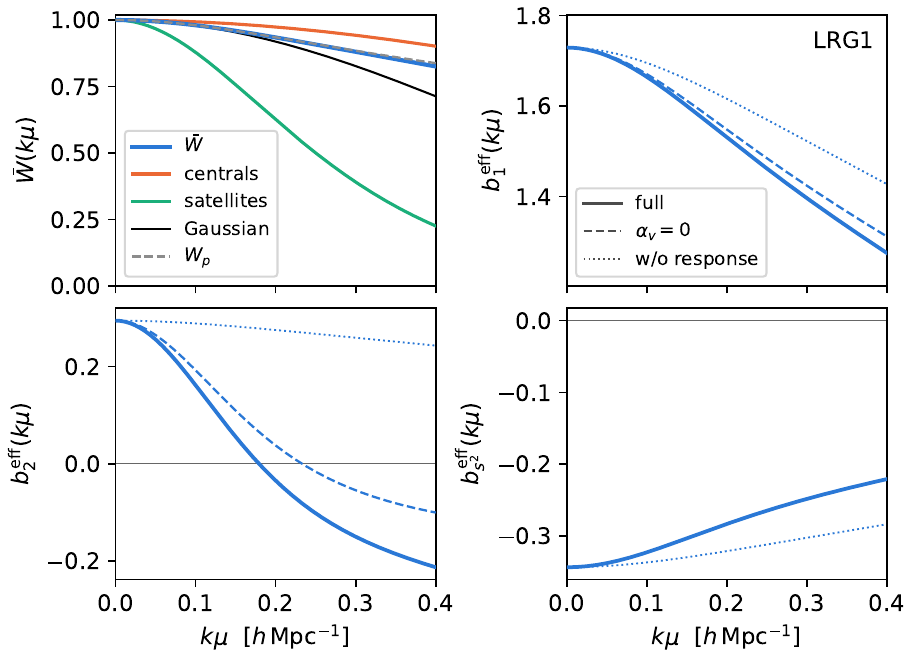}
    \caption{Mean SPCF (top left) and effective bias parameters for $\delta, \delta^2, s^2$ dressed by FoGs, as estimated in the halo model.  For the former, we show contributions from central and satellite galaxies, and compare to the Gaussian and two-parameter $W_p$ forms, with the former a much worse fit despite all constituent galaxies satisfying a Gaussian distribution. For the latter plots we compare the full result (solid) to the mean-field form (dotted), showing that the two behave significantly differently at high $\lambda = k\mu$. For the density biases we in addition show the halo-model result with and without the velocity response $\alpha_v$. }
    \label{fig:damping-biases-lrg1}
\end{figure}

The first panel in Figure~\ref{fig:damping-biases-lrg1} shows the mean damping for the LRG1 sample in DESI when resumming the SPCF of FoGs, as a function of $\lambda = k\mu$. While the individual FoG SPCFs, or damping functions, at each mass are Gaussian, the mean over masses and galaxy types is not, as can be seen by comparing to the Gaussian curve matched to leading order in $\lambda$. This underscores that even if an assumed damping function is true for one population, the same conclusion will not hold in general.

In fact, as can be seen in the same panel, for galaxy samples whose main FoG contribution derives from satellites, a reasonably accurate ansatz for the average damping is\footnote{We thank the LLM Claude for suggesting this ansatz.} 
\begin{equation}
    \bar{W}(\lambda) = \frac{1}{\left(1 + \frac{\lambda^2 \sigma^2_{\rm eff}}{2p}\right)^p }
    \label{eqn:ansatz}
\end{equation}
where, to very good approximation, 
\begin{equation}
    \sigma^2_{\rm eff} = \langle \sigma^2_v \rangle, \quad p = \frac{f_{\rm sat}}{\mathcal{R}_s - f_{\rm sat}}.
\end{equation}
where $\mathcal{R}_s - 1 \approx 1$ is the ratio of the variance of $\sigma_u^2$ across the distribution of satellites to the squared mean. This is due to the fact that the distribution of satellite $\sigma_u^2$'s is very closely related to halo mass function and rises as a power law at low $\sigma_u^2$ (low mass) before an exponential cut-off at large values (high mass), i.e. is well-described by a Gamma distribution. We describe this approximation in more detail in Appendix~\ref{app:stochastic-generator}, where we also compare its agreement with DESI HOD predictions, showing that it significantly out-performs 1-parameter models like Gaussians or Lorentzians.

This simple model illustrates that universal 1-parameter models with a characteristic velocity width only are insufficient to capture the damping responses and means, and that sample characteristics such as the satellite fraction play significant roles. On the other hand, if this 2-parameter model can be shown empirically to match the data well, it may raise the prospect of a simplified response-function formalism based on long-mode derivatives of the mean velocity width and satellite fraction.

The next three panels of Figure~\ref{fig:damping-biases-lrg1} show the effective linear, quadratic, and tidal biases of the LRG1 sample when FoG effects are resummed, as a function of $\lambda = k\mu$. The $\lambda$-dependent coefficient of each bias operator exhibits notably different behavior, starting with the linear bias $b_1$. Notably, in this toy model, the FoG-resummed quadratic bias has a zero crossing at $k\mu \approx 0.2\kMpc$ and deviates significantly from the naive, response-free damping shown in the dotted line. This is because both $b^L_2$ and $\sigma_u(M)$ scale strongly with the halo mass, such that the halos contributing most to the quadratic bias also lead to the most FoG damping. Indeed, abundance-weighting is the leading effect responsible for this shape change, as can be seen by comparing to the dashed line where the response of $\sigma_u(M)$ is set to zero. The tidal-shear bias shows a similar effect with the reverse sign, since it becomes steadily more negative with increasing mass.

\begin{figure}
    \centering
    \includegraphics[width=\linewidth]{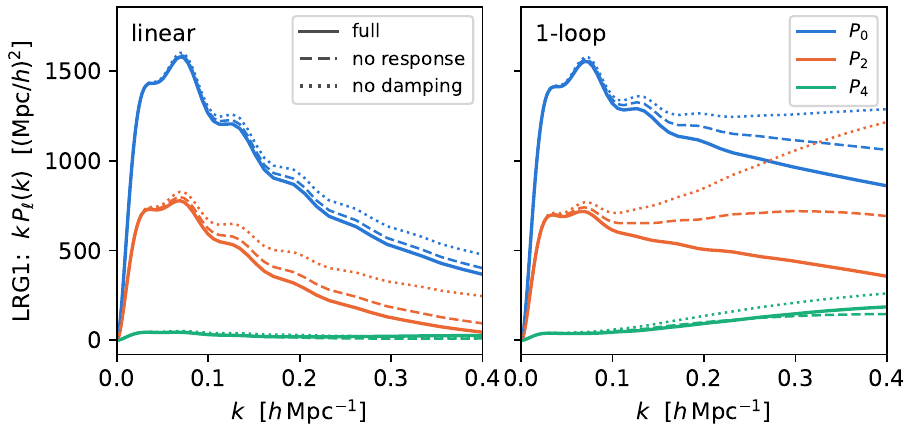}
    \caption{Linear (left) and 1-loop (right) power spectrum multipoles taking into account FoG effects estimated from the halo model for the DESI LRG1 sample. Solid lines show the full calculation including long-wavelength responses, while dashed lines show the mean-field damping result and dotted lines show the un-damped EFT predictions. Dropping the environmental response produces corrections of comparable magnitude to dropping the mean-field damping.}
    \label{fig:pk_lrg1}
\end{figure}

In order to demonstrate the effect of FoGs on practical analyses, Figure~\ref{fig:pk_lrg1} shows the linear and 1-loop power spectra of the LRG1 without FoG resummation, with a single response-free damping function $W$, and in the full formalism. We compute these power spectra using \texttt{velocileptors} \cite{Chen2020,Chen2021b}, and independently with a modified version of \texttt{Folps} \cite{Aviles:2021que,Noriega:2022nhf}, which we modified to allow for bias coefficients that depend on $k\mu$, and set all counterterms and stochastic parameters to zero. In both cases, the changes to the quadrupole between the no-damping and damping predictions are comparable to the changes when neglecting the environmental dependence of the SPCF. Indeed, the changes at 1-loop are particularly noticeable, where it is clear that neglecting response terms in favor of a single damping function can produce order one errors at small scales.

\begin{figure}
    \centering
    \includegraphics[width=\linewidth]{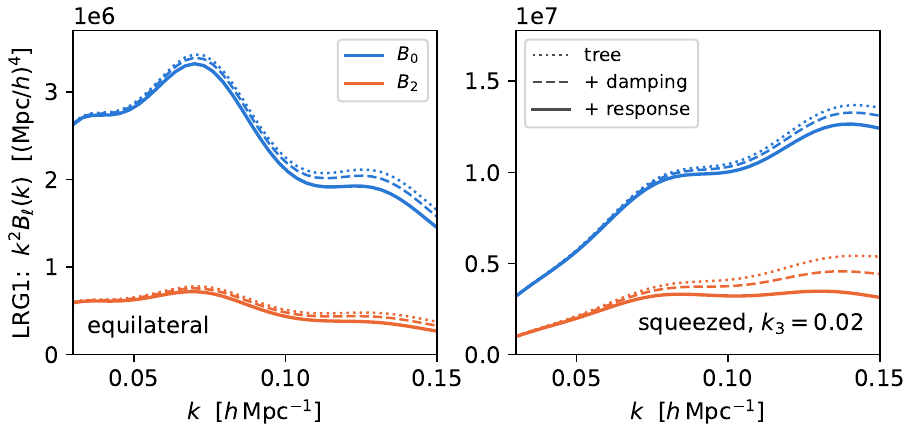}
    \caption{Same as Figure~\ref{fig:pk_lrg1} but for the bispectrum monopole and quadrupole. The left and right panels show equilateral and squeezed triangles, respectively.}
    \label{fig:bk_lrg1}
\end{figure}

Similarly, Figure~\ref{fig:bk_lrg1} shows the bispectrum monopole and quadrupole for equilateral and squeezed---defined as $B_\ell(k,k,k_{\rm min})$ for $k_{\rm min} = 0.02 \kMpc$---at tree level, with the mean damping, and with long-wavelength responses included. The breakdown of the average-damping-only prescription is particularly severe here, with the monopole differing by around $10\%$ and the quadrupole differing by up to $30\%$ by $k = 0.15 \kMpc$, reflecting the steep mass scaling of the quadratic biases. Indeed, this sharp departure from tree level or simple damping factors at relatively large scales suggests one reason that fits to the bispectrum may be more sensitive to scale cuts than the power spectrum is that higher-order bias coefficients, which are leading order in the bispectrum but next-to-leading in the power spectrum, are more sensitive to FoG effects, as we saw in Figure~\ref{fig:damping-biases-lrg1}.

\subsection{A Recipe for Anisotropic Stochasticity: Can $P_{\rm stoch}$ be resummed?}

We end this section on a speculative note. At 1-loop in perturbation theory, the galaxy stochastic power spectrum in redshift space is given by
\begin{equation}
    P_{\rm stoch}(\vec k) = \frac{1}{\bar{n}} \left( 1 + P_{\rm shot} + a^{\rm fog}_{01} (k\mu)^2 \right).
\end{equation}
As noted in Section~\ref{ssec:cts}, going beyond perturbative scales and re-summing FoG contributions requires promoting the coefficients to functions of $\lambda$, and lies outside the scope of this work. However, it is nonetheless interesting to ask if we can make any phenomenological observations about the leading contributions. In particular, the anisotropic coefficient, $a^{\rm fog}_{01}$, is a standard ingredient of fits to galaxy clustering statistics, and has been robustly detected in both simulated halos and galaxies \cite{Schmittfull2021,Ivanov2025}. A curious feature of these detections is that the sign of $a^{\rm fog}_{01}$ tends to be positive: if, for example, the Poisson noise were to be damped by FoGs, one might expect that $a^{\rm fog}_{01}$ be negative of order $\sigma_u^2/\nbar$. However, within the formalism we have developed above, it is clear that any such damping could only apply to $P_{\rm shot}$, i.e. deviations from Poisson-noise due to correlations at short scales from distinct pairs. For massive samples, a leading effect is halo exclusion, which limits halos from overlapping, creating a sub-Poisson signal \cite{Baldauf2013}. The damping of this negative $P_{\rm shot}$ then produces a positive anisotropy along the line of sight: redshift space distortions remove exclusions because random virial motions map objects into otherwise forbidden exclusion zones. On the other hand, the same logic predicts that super-Poisson samples, i.e. those with high satellite fractions, will have a negative $a^{\rm fog}_{01}$. Indeed, this is precisely what was observed in ref.~\cite{Ivanov2025} for realistic simulations of galaxies. Within the halo model, it is possible to at least crudely estimate both of these stochastic effects in a similar way as we have estimated the deterministic responses---without accounting for intra-halo density-velocity correlations, this can be simply accomplished by multiplying the halo profile $u(k|M)$ by the SPCF (see e.g. ref.~\cite{Baldauf2013} for the real-space counterpart of these calculations). However, in general it would be preferable to link the halo density and velocity profiles, which are connected under equilibrium, leading to a non-separable redshift-space halo profile---we leave further exploration of these effects to future work.

\section{Conclusions}
\label{sec:conclusions}

The large physical scale associated with the nonlinear redshift-space mapping, imparted by the small-scale, virial motions of galaxies, is one of the main obstacles to extending the perturbative reach of effective field theory (EFT) predictions of galaxy clustering in spectroscopic surveys. Since these small-scale velocities live in the ultraviolet regime, their precise properties cannot be inferred from the EFT, where their impact on galaxy clustering is absorbed into counterterms at each order, but rather may be inferred by observations or simulations of small-scale astrophysics. 
The primary effect of these velocities is a smearing along the line of sight governed by the length of the wavenumber along the line of sight $k\mu$ and the characteristic length $\sigma_u$ of the redshift-space mapping induced by small-scale motions. At $k\mu < \sigma_{u}$, the contributions of these velocities  can be captured by EFT counterterms, but their presence limits the reach of the EFT, which otherwise is perturbative up to $k_{\rm nl} \gg \sigma_{u}^{-1}$. This observation has led to many empirical prescriptions for the so-called ``Fingers of God'', i.e. the impact of these small-scale velocities on redshift-space galaxy clustering, based on empirical models where e.g. virial motions are distributed via Maxwellian or exponential distributions.  Beyond perturbative modeling, these velocity distributions can further be connected to galaxy formation and the halo occupation distributions of galaxies, leading to a rich phenomenology.

Our goal in this work has been to explore whether it is possible to non-perturbatively capture, or re-sum, the effects of FoGs on galaxy clustering, thereby enlarging the reach of EFT predictions of galaxy clustering. Our starting point is the by now well-known observation that the effect of FoGs can be directly linked to the characteristic function of the underlying velocities, i.e. the 2-point function to the 2-particle velocity joint characteristic function, the 3-point function to the 3-particle version, etc. Working in this language, we show that purely long-wavelength contributions to galaxy clustering are dressed by the single-particle characteristic function (SPCF) of their velocity $W(\lambda|\bx)$, and that a similar simplification can be obtained for contributions related to stochastic noise in the independent draw approximation, exact for perfectly relaxed halos, where the intrahalo velocities of galaxies are uncorrelated. Under the mean-field ansatz, where $W(\lambda | \bx)$ is replaced by its spatial average $\bar{W}(\lambda)$, we recover the standard effective damping form in the literature, where the computed power spectrum is multiplied by damping factor $\bar{W}^2$, and similarly for $N$-point functions, with the caveat that Poisson-noise contributions are un-damped.

Expanding in the environmental dependence of $W(\lambda|\bx)$, we show that the FoG-dressed galaxy density field admits an expanded bias expansion due to the line-of-sight dependence of the velocity distribution. At leading order in the gradient expansion, these bias operators are precisely the selection-effect expanded basis of ref.~\cite{Desjacques2018}. At higher order in derivatives, however, this expansion also admits parity odd operators, since the argument of the SPCF $\lambda = k\mu$ is odd. We derive selection rules for these operators, and their low-$k$ limits $\lambda^N$, based on symmetries and the form of the redshift-space mapping. The long-wavelength responses of $W(\lambda|\bx)$ modify the perturbation-theory kernels of the galaxy density with coefficients that are functions of $\lambda$; consequently, unlike in the mean field limit, the FoG effect on different terms in the perturbative expansion cannot be re-summed via a single function $\bar{W}(\lambda)$.

Finally, we perform estimates for these long-wavelength responses of FoG damping within the halo model. Physically, these responses appear because the presence of long-wavelength modes changes (a) the mass, and therefore velocity-dispersion, distribution of halos and (b) the intra-halo dynamics of each halo. We show that the sizes of these responses for DESI-like galaxies are comparable to those from the commonly-used mean-field correction itself, with the discrepancy worsening for higher-order operators, suggesting that consistently resumming FoG effects requires this large expansion of degrees of freedom in the galaxy bias expansion. We show that the discrepancies persist at the level of $N$-point functions like the power spectrum and bispectrum, which we compute with and without the responses.

We have thus found that the impact of FoGs on the deterministic response of the redshift-space galaxy density \textit{may} be resummed, but at the cost of significantly expanding the bias basis, both because the breaking of rotational symmetry by the line of sight increases the number of allowed operators, and because each operator obtains a coefficient that is a free function of $\lambda = k\mu$. Importantly, a resummation that captures all numerically-large FoG contributions must be significantly more complicated than the mean-field ansatz where galaxy statistics are simply damped by a (often simply parametrized) function. On the other hand, since all FoG-induced effects can be subsumed into functions \textit{strictly} of $\lambda$, it may be that non-parametrically implementing these functions and including them in galaxy clustering analyses is a robust way to formulate the theoretical covariance associated with FoGs from the model side, complementing data-driven approaches such as the $Q_0$ statistic \cite{Ivanov2022,DAmico2024} or approaches at the level of sample construction \cite{Baleato2025}.

Several aspects of the phenomenology of FoGs outlined in this work merit further investigation. First, while we have argued that FoGs respond anisotropically to their environment, in this work we have only obtained estimates for the isotropic contributions using the halo model. We have argued that the responses should be small---amounting to intrinsic alignments of the velocity distributions of halos, similar to those of their shapes---but it should be possible to directly measure them, for example by measuring the dynamics of dark matter particles of halos in tidal separate-universe simulations which simulate the formation of structure under the influence of a long-wavelength tidal field \cite{Shogo2020,Stucker2021,Akitsu2023b}. In this approach, the scale-dependence due to FoGs alone can be isolated, providing a further handle on the size of pure FoG corrections compared to other nonlinearities. Alternatively, these responses may be measurable directly by measuring the scale-dependent bias of anisotropic EFT operators at the field level, as has been done within this basis for the Lyman-$\alpha$ forest \cite{Cieplak2016,Belsunce2025,Belsunce2026a,Belsunce2026b}, who also use it to elucidate the structure of the stochastic sector.

\section*{Acknowledgements}

We thank Mark Maus for helpful discussions in the early stages of this project. We thank Martin White, Zvonimir Vlah,  Misha Ivanov, Gustavo Niz, Davide Bianchi, Mathias Garny, Alex Eggemeier and Jiamin Hou for useful conversations. Support for this work was provided by NASA through the NASA Hubble Fellowship grant HST-HF2-51572.001 awarded by the Space Telescope Science Institute, which is operated by the Association of Universities for Research in Astronomy, Inc., for NASA, under contract NAS5-26555. AA is supported by SECIHTI grants CBF2023-2024-162 and CBF-2025-I-2795, and by DGAPA-PAPIIT IA101825. This work was performed in part on the tarmac at Syracuse Hancock International Airport.

\appendix
\section{Stochastic realizations of the Single-Particle Characteristic Function}
\label{app:stochastic-generator}

\subsection{Non-Gaussian Realizations of the SPCF}

The large-separation factorization in Equation~\ref{eqn:long_factorization} identifies the SPCF $W(\lambda)$ as the elementary FoG object. Here we give simple stochastic realizations of this SPCF and then connect them to the galaxy model of Equation~\ref{eqn:ansatz}.
For a Gaussian residual velocity we may write
\begin{equation}
    u_{\parallel}^{\rm FoG} = \sigma_\text{FoG} Y, \qquad Y\sim\mathcal N(0,1),
    \label{eqn:app-gaussian-process}
\end{equation}
Here $\sigma_\text{ FoG}^2\equiv\avg{(u_{\parallel}^\text{FoG})^2}$  characterizes the unresolved residual velocity itself, and bulk flows are assigned to $\bu_\text{PT}$. 
The corresponding SPCF is the Gaussian form
\begin{equation}
W_\text{Gaussian}(\lambda)=\exp\left(-\frac{1}{2}\lambda^2\sigma_\text{FoG}^2\right).
\label{eqn:app-gaussian-generator}
\end{equation}

A simple stochastic realization of this Gaussian-mixture mechanism is obtained by modulating one Gaussian variable by another,
\begin{equation}
u_{\parallel}^{\rm FoG}=\sigma_\text{FoG}XY, \qquad X,Y\sim\mathcal N(0,1),
\label{eqn:app-xy-process}
\end{equation}
with $X$ and $Y$ independent. At fixed $X$, the velocity is Gaussian with variance $\sigma_\text{FoG}^2X^2$, and averaging over $X$ gives
\begin{equation}
W_\text{Lorentz}(\lambda) = \avg{\exp\left(-\frac{1}{2}\lambda^2\sigma_\text{FoG}^2X^2\right)}_X = \frac{1}{\sqrt{1+\lambda^2\sigma_\text{FoG}^2}},
\label{eqn:app-xy-generator}
\end{equation}
This is the square root of the conventional pairwise Lorentzian damping $(1+\lambda^2\sigma_{\rm FoG}^2)^{-1}$, corresponding to an exponential pairwise velocity distribution, with related Gaussian-mixture constructions discussed in \cite{Sheth1996,Bianchi2016}. Equivalently, defining $S=\sigma_\text{FoG}^2X^2$ gives
\begin{equation}
u_{\parallel}^{\rm FoG}=\sqrt{S}\,Y, \qquad S\sim \mathrm{Gamma}\!\left(\frac{1}{2},2\sigma_\text{FoG}^2\right),
\label{eqn:app-xy-gamma}
\end{equation}
so Equation~\ref{eqn:app-xy-generator} is the Laplace transform of a fluctuating local Gaussian variance. 

A third example is obtained by allowing the Gaussian modulation to have a nonzero mean, providing a noncentral extension of the Gamma-variance construction above,
\begin{equation}
u_{\parallel}^{\rm FoG}=(\sigma_0+ \sigma_1 X)Y,
\qquad
X,Y\sim{\cal N}(0,1),
\end{equation}
with $X$ and $Y$ independent, we obtain the SPCF
\begin{equation}
W_\text{non-central}(\lambda)= \frac{1}{(1+\sigma_1 ^2\lambda^2)^{1/2}} \exp\left[ -\frac{\sigma_0^2\lambda^2}{2(1+\sigma_1 ^2\lambda^2)} \right].
\end{equation}

In the virialized small-scale limit, where the velocity statistics are dominated by the distribution of local velocity dispersions, the noncentral construction above, when applied to the pairwise line-of-sight velocity difference rather than to the single-particle velocity, yields the functional form found in \cite{Bianchi2016} and later adopted in the VDG model \cite{Sanchez2017,Eggemeier2025}. A related halo-mixture construction in \cite{Sheth1996} shows how averaging Gaussian intrahalo velocities over the halo population can generate an approximately exponential pairwise velocity distribution.

The previous examples demonstrate that statistical mixtures of SPCFs with simple forms, in this case Gaussians, can generate a wide variety of functional forms not well-described by the original. Intuitively, this is because averaging over a mixture of Gaussians with different width blends populations with very different scale dependences, in this case the width of the Gaussian, where e.g. a wide Gaussian can stay constant when a narrow one, part of the same average, has effectively decayed to zero. For realistic galaxy samples, this kind of mixture can come about both from the dependence on host halo mass as well as from distinct populations like centrals and satellites. As shown in Figure~\ref{fig:damping-biases-lrg1}, this suggests that SPCFs of realistic galaxy samples cannot, even for minimal HOD descriptions, be reduced to a simple rescalings of 1-parameter functional forms. In the subsection below, we investigate whether we can nonetheless make some progress with a 2-parameter functional form.

\subsection{Galaxy mixtures and the $W_p$ model}
\label{app:wp-galaxies}

\begin{figure}
    \centering
    \includegraphics[width=\linewidth]{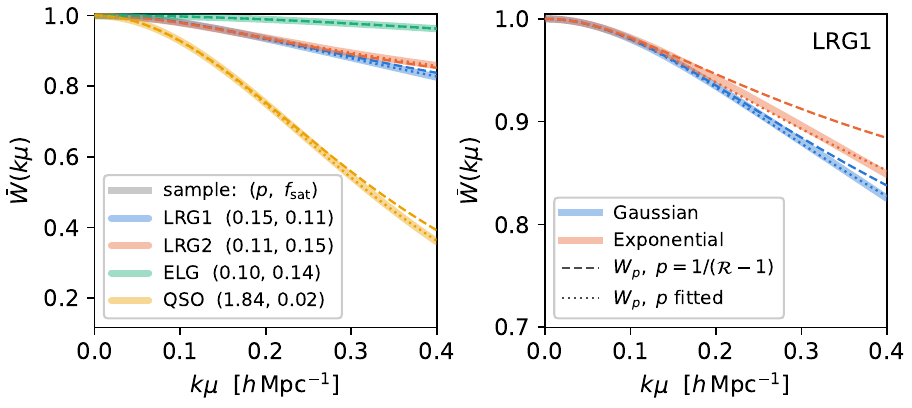}
    \caption{(Left) Comparison of the mean SPCF of HODs of four DESI galaxy samples with the 2-parameter $W_p$ ansatz. In all cases, $W_p$ gives an excellent match to the mean damping for $p$ in close agreement with the satellite fraction, as expected, with the exception of the QSO HOD which possesses an anomalous central velocity bias, likely to compensate for redshift errors. (Right) HOD prediction for LRG1 when intra-halo velocities are described by Gaussian or Lorentzian profiles corresponding to exponential distributions, compared to the $W_p$ model. In this case it is necessary to tune $p$ to fit the mean SPCF rather than obtain it directly from moment matching. }
    \label{fig:wp}
\end{figure}

For the minimal central--satellite model used in the main text, we take centrals to have no residual velocity with respect to the halo bulk motion, $u_{\parallel,\text{c}}^{\rm FoG}=0$, and hence $W_\text{c}(\lambda)=1$. If $f_\text{sat}$ is the satellite fraction and $W_\text{s}$ is the satellite-averaged SPCF
then
\begin{equation}
W_\text{HOD}(\lambda)=1-f_\text{sat}+f_\text{sat}W_\text{s}(\lambda).
\label{eqn:app-whod}
\end{equation}
Let $S=\sigma_\text{sat}^2$ denote the residual FoG velocity variance associated with a tracer draw's satellite, while the variance for centrals is zero. We assume a conditional Gaussianity for the satellite unresolved velocity,
\begin{equation}
u_{\parallel,\text{s}}^{\rm FoG}=\sqrt{S}\,Y,\qquad Y\sim\mathcal N(0,1),
\end{equation}
with the variance $S$ varying across the population. Hence, averaging over the Gaussian variable $Y$ we obtain
\begin{equation}
W_\text{s}(\lambda)=\avg{e^{-\lambda^2 S/2}}_\text{s}.
\label{eqn:app-wsat}
\end{equation}
We can characterize the satellite distribution of variances by the ratio
\begin{equation}
\mathcal{R}_s\equiv\frac{\avg{S^2}_\text{s}}{\avg{S}_\text{s}^{2}} = 1 + \frac{\avg{S^2}_\text{s}-\avg{S}_\text{s}^{2}}{\avg{S}_\text{s}^{2}}
\label{eqn:app-rs}
\end{equation}
such that $\mathcal{R}_s-1$ measures the departure from a population in which all satellites sample the same velocity variance.
For the full tracer sample we have (where we now use $\avg{\cdots}\equiv \avg{\cdots}_\text{HOD}$)
\begin{equation}
\avg{S}=f_\text{sat}\avg{S}_\text{s},
\label{eq:S1}
\end{equation}
and
\begin{equation}
\avg{S^2}=f_\text{sat}\avg{S^2}_\text{s}=f_\text{sat}\mathcal{R}_s\avg{S}_\text{s}^{\,2}=\frac{\mathcal{R}_s}{f_\text{sat}}\avg{S}^{2}.
\label{eq:S2}
\end{equation}

Equivalently, the exact variance distribution of the minimal central--satellite model is
\begin{equation}
P_\text{HOD}(S)=(1-f_\text{sat})\delta_D(S)+f_\text{sat}P_\text{s}(S),
\label{eqn:app-phod}
\end{equation}
where $P_\text{s}(S)$ is the satellite-velocities variance distribution.  Now, the Lorentzian example of the previous subsection suggests a simple effective representation of $P_\text{HOD}(S)$. There, the fluctuating variance follows a Gamma distribution with fixed shape $p=1/2$, and its Laplace transform gives the Lorentzian SPCF. For galaxies, the central--satellite mixture instead changes both the mean and the spread of the variance distribution. Since the velocity dispersion $S$ is a proxy for halo mass, and the mass function cuts off steeply at high masses while satellite fraction grows roughly as a power law, we can model its distribution as a Gamma distribution
\begin{equation}
P_\text{HOD}^{\Gamma}(S)=\frac{S^{p-1}e^{-S/\theta}}{\Gamma(p)\theta^p}.
\label{eqn:gamma_p}
\end{equation}
In order to match the HOD expectation asymptotically at low $\lambda$ we can then determine its two parameters by matching the first two moments of the tracer distribution given by Equations \ref{eq:S1} and \ref{eq:S2}. That is, from the last equation
we have $\avg{S}=p\theta$ and $\avg{S^2}=p(p+1)\theta^2$, and therefore $\avg{S^2}/\avg{S}^2 = 1+\frac{1}{p}$. By matching moments we obtain
\begin{equation}
p=\frac{f_\text{sat}}{\mathcal{R}_s-f_\text{sat}}.
\label{eqn:app-p-fsat}
\end{equation}
Defining the mean residual variance of the full tracer population as $\sigma_\text{eff}^2\equiv\avg{S}=f_\text{sat}\avg{S}_\text{s}$, the Gamma scale parameter is $\theta = \sigma_\text{eff}^2/p$. The corresponding SPCF is the Laplace transform of the effective variance distribution,
\begin{equation}
W_p(\lambda)=\int_0^\infty dS\,P_\text{HOD}^{\Gamma}(S)e^{-\lambda^2S/2}=\left(1+\frac{\lambda^2\sigma_\text{eff}^2}{2p}\right)^{-p}.
\label{eqn:app-wp}
\end{equation}
This is the form used in Equation~\ref{eqn:ansatz}. For samples where centrals contribute meaningfully to the SPCF, it can be useful to instead match $p$ to the joint velocity distribution directly, equivalent to setting $f_{\rm sat} = 1$ in Equation~\ref{eqn:app-p-fsat} and replacing $\mathcal R_s$ with the joint central and satellite moment ratio $\mathcal R$.

Figure~\ref{fig:wp} compares the Gamma-distribution ansatz to HODs employed by the DESI collaboration to LRGs, ELGs and QSOs \cite{Rocher2023,Yuan2024}. In the left panel we compare the mean-field damping $\bar{W}$ of each of the four DESI samples versus the Gamma-function ansatz. The dashed lines show the result when $p$ is directly predicted from matching the moments of $S$ in the HOD prescription. As tabulated in the legend, these values are remarkably close to the satellite fraction $f_{\rm sat}$, since $\mathcal{R}_s$ is an order-one number. The only large deviation between $p$ and $f_{\rm sat}$ occurs in the case of the QSOs---this is because the HOD chosen by DESI assigned a large velocity bias to \textit{centrals}, possibly due to mis-calibration of observational redshift errors \cite{Yuan2024,Yu2024}, violating the assumptions of the derivation. The QSO case thus provides a useful example of where the simple relation between $p$ and $f_{\rm sat}$ breaks down, allowing us to test the robustness of the parametrization. Similarly, but to a much lesser extent, the HOD of LRG1 also features a small but non-zero value of the velocity bias for centrals, leading to $p$ larger than $f_{\rm sat}$. The dotted lines further show the agreement of the $W_p$ form with the full HOD calculation when $p$ is fitted to minimize the error between $0.1-0.4 \kMpc$. The Gamma distribution ansatz is thus an excellent one for all four shown HODs, though $p$ only has the interpretation of satellite fraction for appropriately physical models of centrals and satellites. In order to further demonstrate the versatility of the Gamma distribution ansatz the right panel explores the case where the intra-halo velocity distribution is given by not a Gaussian (blue), but is exponential (i.e. the characteristic function in Lorentzian). In this case, the Gamma function ansatz is no longer valid; nonetheless, by freeing $p$, we are able to obtain percent-level agreement out to $0.4 \kMpc$.

\bibliographystyle{JHEP}
\bibliography{main}
\end{document}